# Wafer-scale detachable monocrystalline Germanium nanomembranes for the growth of III-V materials and substrate reuse

Nicolas Paupy[1,2], Zakaria Oulad Elhmaidi[1,2], Alexandre Chapotot[1,2], Tadeáš Hanuš[1,2], Javier Arias-Zapata[1,2], Bouraoui Ilahi[1,2], Alexandre Heintz[1,2], Alex Brice Poungoué Mbeunmi[1,2], Roxana Arvinte[1,2], Mohammad Reza Aziziyan[1,2], Valentin Daniel[1,2], Gwenaëlle Hamon[1,2], Jérémie Chrétien[1,2], Firas Zouaghi[1,2], Ahmed Ayari[1,2], Laurie Mouchel[1,2], Jonathan Henriques[1,2], Loïc Demoulin[1,2], Thierno Mamoudou Diallo[1,2], Philippe-Olivier Provost[1,2], Hubert Pelletier[1,2], Maïté Volatier[1,2], Rufi Kurstjens[3], Jinyoun Cho[3], Guillaume Courtois[3], Kristof Dessein[3], Sébastien Arcand[4], Christian Dubuc[4], Abdelatif Jaouad[1,2], Nicolas Quaegebeur[5], Ryan Gosselin[6], Denis Machon[2,7], Richard Arès[1,2], Maxime Darnon[1,2] & Abderraouf Boucherif[1,2*]

1- *Institut Interdisciplinaire d'Innovation Technologique (3IT), Université de Sherbrooke, 3000 Boulevard de l'Université, Sherbrooke, J1K 0A5, QC, Canada*
2- *Laboratoire Nanotechnologies Nanosystèmes (LN2) - CNRS IRL-3463, Université de Sherbrooke, 3000 Boulevard Université, Sherbrooke, Québec J1K OA5, Canada*
3- *Umicore Electro-Optic Materials, Watertorenstraat 33, 2250, Olen, Belgium*
4- *Saint-Augustin Canada Electric Inc.*
   *75 rue d'Anvers, Saint-Augustin, G3A 1S5, QC, Canada*
5- *Department of Mechanical Engineering, Université de Sherbrooke, 2500 Boulevard de l'Université, Sherbrooke, J1K 2R1 QC, Canada*
6- *Department of Chemical and Biotechnological Engineering, Université de Sherbrooke, 2500 Boulevard de l'Université, Sherbrooke, J1K OA5, QC, Canada*
7- *Institut Lumière Matière, UMR5306 Université Lyon 1-CNRS, Université de Lyon F-69622 Villeurbanne cedex, France*

*Corresponding author. Email: abderraouf.boucherif@usherbrooke.ca

Germanium (Ge) is increasingly used as a substrate for high-performance optoelectronic, photovoltaic, and electronic devices. These devices are usually grown on thick and rigid Ge substrates manufactured by classical wafering techniques. Nanomembranes (NMs) provide an alternative to this approach while offering wafer-scale lateral dimensions, weight reduction, limitation of waste, and cost effectiveness. Herein, we introduce the Porous germanium Efficient Epitaxial LayEr Release (PEELER) process, which consists of the fabrication of wafer-scale detachable monocrystalline Ge NMs on porous Ge (PGe) and substrate reuse. We demonstrate monocrystalline Ge NMs with surface roughness below 1 nm on top of nanoengineered void layer enabling layer detachment. Furthermore, these Ge NMs exhibit compatibility with the growth of III-V materials. High-resolution transmission electron microscopy (HRTEM) characterization shows Ge NMs crystallinity and high-resolution X-ray diffraction (HRXRD) reciprocal space mapping endorses high-quality GaAs layers. Finally, we demonstrate the chemical reconditioning process of the Ge substrate, allowing its reuse, to produce multiple free-standing NMs from a single parent wafer. The PEELER process significantly reduces the consumption of Ge during the fabrication process which paves the way for a new generation of low-cost flexible optoelectronics devices.

## 1. Introduction

Free-standing semiconductor nanomembranes (NMs) draw tremendous attention in nanoscience and engineering for their unique mechanical stability and structural properties.[1] In this regard, they represent a powerful technology for monocrystalline growth, layer release, and transfer to any target substrate.[2] Besides, this scheme allows significant substrate material reduction compared to conventional substrates manufactured by classical wafering techniques[3,4]. This is especially interesting in the case of expensive materials.[5] Furthermore, thin NMs enable the improvement of semiconductor properties and the manufacture of novel devices such as memories and sensors,[6] high-performance photodetectors,[7] and photovoltaic devices.[8]

Ge NMs attract special attention since the intrinsic properties of Ge grants the growth compatibility with a wide range of III-V alloy materials.[9,10] Moreover, Ge NMs benefit from flexibility and lightweight for the fabrication of high efficiency and flexible optoelectronic devices.[11] Additionally, vertical miniaturization of optoelectronic devices was recently shown using Ge NMs.[12,13] Some challenges are yet to be overcome to achieve widespread commercial adoption of Ge NMs including (a) a cost-effective and viable fabrication process; (b) wafer-scale application (c) compatibility with the semiconductor manufacturing procedures.

Ge NMs and substrate reuse offer a solution to reduce the overall consumption of Ge, which is a critical raw material. Moreover, the use of Ge NMs would greatly reduce the weight of III-V triple-junction solar cells on Ge substrate, which represents more than 90% [14] of the total mass of the solar cell. The use of NMs would also allow the reduction of the Ge substrate thickness (below 60 um[14]), which has been shown to be a key technology for the next generation of space and some terrestrial applications, like electric cars, drones or wearable electronic devices.

This strategy further complements the on-going efforts to develop a sustainable process flow from Ge extraction to device processing.[13] Different techniques for detachable layers and substrate reuse have been proposed: Epitaxial lift-off of layers (ELO) using a selective chemical etching of a sacrificial layer to release the "active" epitaxial layer.[15] ELO is the most widespread technique proposed in the literature. However, at wafer-scale the ELO process copes with some challenges, such as thin-film handling and alignment for device fabrication.[16] Another technique that has been the subject of recent research for : controlled spalling[17]. It has been shown that the substrate can be reused, but this technique generates many defects, which will strongly impact the performance of the devices.

Recently, another technique called Germanium-on-Nothing (GoN) uses a sequence of lithography, plasma etching, and annealing steps to create a weak void layer used for the detachment. However, this approach increases the substrate reuse cost and the complexity of the process.[18] Other layer detachment and substrate reuse methods were demonstrated such as laser lift-off,[19] Ge-on-Ge lift-off[20] and double-layer porosification lift-off.[21]
Nonetheless, none of the preceding techniques have successfully demonstrated all the following points necessary for a wide adoption of the method: (1) fabrication of detachable and mono-crystalline Ge NMs (2) wafer-scale application (3) cost effective reconditioning of the substrate after NMs detachment (4) Multiple re-uses of the parent substrate.

In the present work, we introduce the Porous germanium Efficient Epitaxial LayEr Release (PEELER) process for wafer-scale growth of monocrystalline Ge NMs and their detachment, compatible with Ge substrates reuse. This approach consists of four key steps: (i) Wafer-scale porosification by bipolar electrochemical etching (BEE) of a 100 mm, 6° off-cut Ge wafer (ii) Growth of a monocrystalline Ge on porous Ge (PGe) by molecular beam epitaxy (MBE) (iii) Ge NMs detachment from the substrate (iv) Substrate reconditioning by chemical etching to enable reuse of the substrate and cycle repetition.
Furthermore, we provide detailed microstructural investigations of the epitaxial Ge NMs by High-resolution transmission electron microscopy (HRTEM). The chemical beam epitaxy (CBE) growth of single-phase GaAs on Ge NMs shows the suitability of this type of substrate for the growth of III-V heterostructures. Compared to the traditional epitaxy on conventional

Ge wafers, the analysis of the Ge consumption of the PEELER process testifies considerable reduction of the total Ge material used for the production of high-performance devices.

## 2. Results and Discussion

### Approach

The PEELER approach is illustrated by four steps as shown in the **Figure 1**. The first step (i) involves forming a single PGe layer on a 100 mm Ge substrate with 6° misorientation toward (111). (ii) After chemical cleaning and low temperature annealing (LT), Ge buffer layer has been first deposited at LT followed by high-temperature (HT) annealing step to reorganize the porous layer into weak voided layer below the Ge buffer. Then, a high-quality Ge layer is grown on top to form the Ge NM at higher temperatures. (iii) This Ge NM is detached mechanically from the substrate (iv) Finally, the parent substrate is reconditioned by an optimized chemical polishing and the process is repeated to produce more NMs.

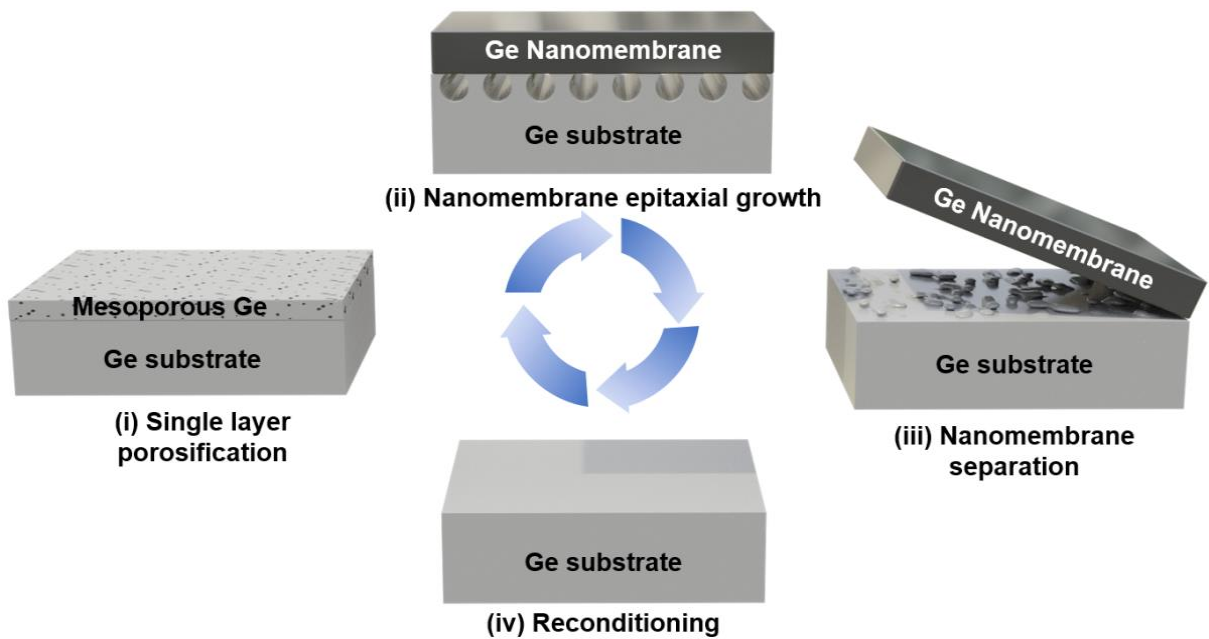

**Figure 1**. Schematic illustration of Ge NMs fabrication and substrate reuse process by the PEELER approach.

### Epitaxial Growth of Ge NMs on PGe

The PEELER cycle starts with the formation of a uniform PGe layer by bipolar electrochemical etching (BEE). The cross-sectional SEM view (Figure 2a) shows a well-defined interface between the PGe layer with sponge-like morphology and the bulk Ge material. Non-destructive ellipsometry mapping with 49 measurement points was used to determine the PGe layer characteristics and to evaluate its uniformity over the entire 100 mm wafer (Figure 2b and 2c). The mean thickness of the PGe layer is 174 ± 3 nm (Figure 2b) which is consistent with SEM image analysis (Figure 2a). In terms of porosity, the mean value is 42 ± 1% (Figure 2c). These results demonstrate excellent uniformity with a standard deviation below 2% for both thickness and porosity. Moreover, the RMS roughness (σ) of the PGe layer is approximately 2 nm (Figure S1). The medium porosity combined with low surface roughness and good lateral homogeneity make this PGe layer an excellent epitaxy template for Ge growth.

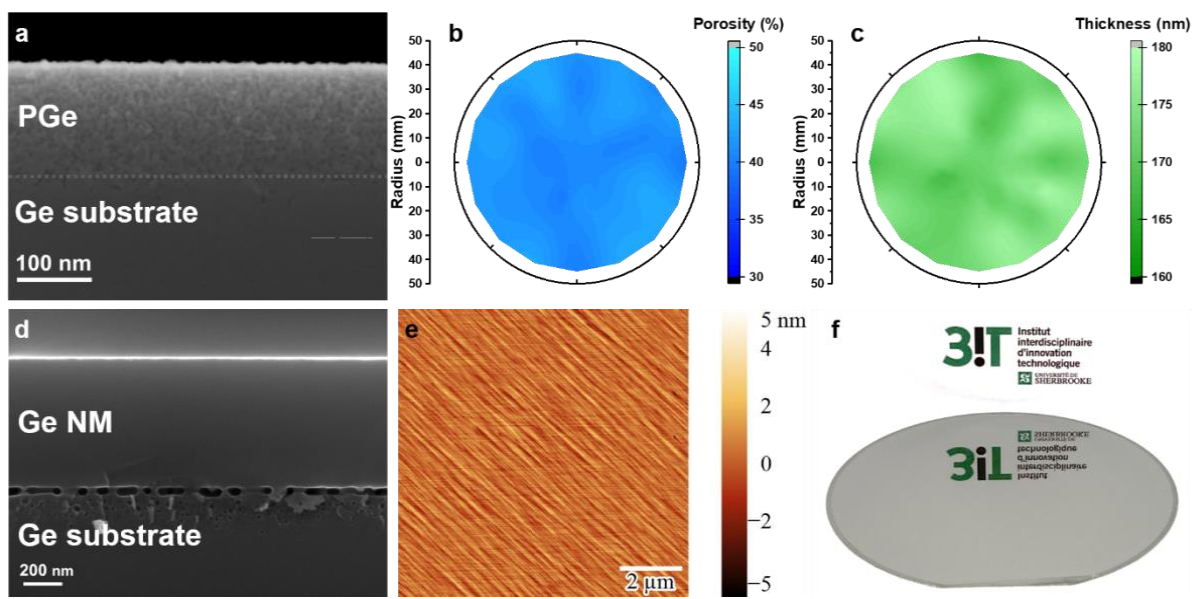

**Figure 2**. (a) Cross-section SEM image of PGe layer. Ellipsometry mapping of the (b) thickness and (c) porosity values of the 100 mm Ge wafer. (d) Cross-section SEM image of the epitaxial Ge layer. (e) AFM image with RMS roughness σ of 0.5 nm for epitaxial Ge/PGe. (f) Image of a typical epitaxial Ge grown on PGe/Ge at wafer-scale, showing that the surface maintains its specular reflection properties (mirror-like finish).

Ge epitaxial growth was carried out on the PGe substrate. The 700 nm thick structure is shown by the SEM cross-sectional view (Figure 2d) is composed of a LT Ge buffer layer and a thicker epitaxial Ge layer. As revealed by the Figure 2d, the porous material morphology has been significantly reorganized during the HT annealing step and the epitaxial growth. With the presence of the buffer layer, the pores of the PGe are closed and the porous material remains confined to the porous/buffer interface during its reorganization. This phenomenon is based on the Ostwald ripening and Rayleigh instability.[22,23] This reorganization allows the formation of a void layer, which facilitates the separation of the Ge epilayer from the substrate; also known as weak separation layer. The presence of pillars in the weak layer allows the mechanical stability of the epitaxial structure and the subsequent lift-off of the Ge NMs. Similar shape transformation of PGe into a voided region has been reported for other porous structures such as porous Si,[24,25] porous InP [26] and porous GaN.[27,28] To investigate the surface quality of the Ge epi-layer, AFM surface analysis (Figure 2e) shows a smooth surface with low roughness of 0.7 nm very close to that of the parent Ge epi-ready substrate (RMS ~ 0.3 nm). A visual inspection of the 100 mm Ge structure shows a mirror-like surface, as shown in Figure 2f.

To determine the crystal quality of the epitaxial structure, the Germanium NMs was analyzed by high-resolution transmission electron microscopy (HRTEM) along the zone axis [110]. The cross-section of the Ge structure (Figure 3a) shows three regions: the epitaxial Ge structure, the weak layer and the Ge substrate. No defects or dislocations are observed in the Ge epi-layer. Selected area electron diffraction (SAED) pattern obtained from Ge epi-layer (Inset of

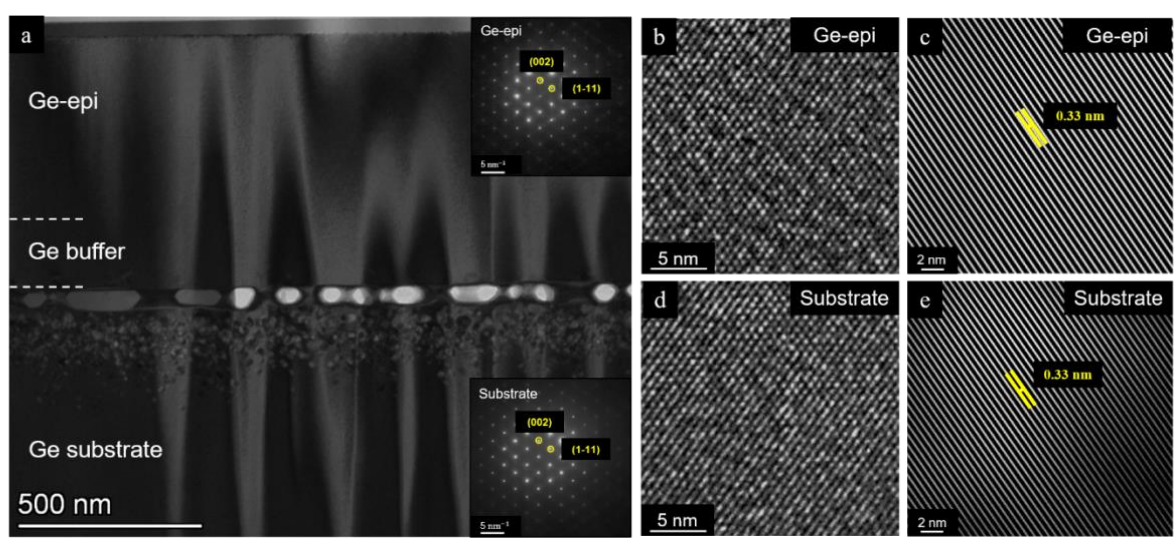

**Figure 3**. (a) Cross section TEM image of the epi-layer Ge structure. (b) HRTEM of Ge epi-layer. Insets in (a) are the selected-area electron diffraction (SAED) patterns from Ge epi-layer and substrate, taken along [110] zone axis. (c) The inverse fast Fourier transform (IFFT) from Ge epi-layer showing only (1–11) crystalline planes. (d) HRTEM from Ge substrate. (e) IFFT from Ge bulk displaying only (1–11) crystalline plane

Figure 3a) shows diffraction spots which demonstrate a monocrystalline pattern [10]. Moreover, the Ge epi-layer pattern is identical to that of the substrate (Inset of Figure 3a) which suggests that the crystallographic information was transmitted from the substrate to the Ge epitaxial structure through epitaxy. This preservation of the crystalline quality is also supported by the similarity of the orientation of the crystallographic planes, shown by the high-resolution zoomed images of the epitaxial layer and the substrate (Figure 3b and 3d).

Fourier mask filtering tools and the inverse fast Fourier transform (IFFT) were applied to determine the interplanar spacing and verify if the 6º off-cut was preserved in the NMs. Figure 3c and 3e show the generated lattice fringes for the Ge epi-layer and Ge bulk, respectively. In these figures, the lattice fringes of the (1–11) planes are indicated by two parallel lines. The interplanar spacing is approximatively 0.33 nm being the same as that of Ge bulk (Figure 3e). This value agrees with the $d_{1-11}$ of the diamond cubic structure of Ge [29]. Therefore, HRTEM observations demonstrate the successful growth of single-crystal Ge on PGe/Ge substrates with the same 6° off-cut as the Ge substrate.

**Growth of High-Quality GaAs on Ge/PGe Structure**

Aiming to test the viability of using the detachable Ge NMs for the growth of III-V compounds, a 400 nm GaAs layer was grown on the Ge structure. The structural quality of the GaAs/Ge/PGe structure was evaluated by using high-resolution XRD in $\omega - 2\theta$ configuration, around (004) symmetric and (115) asymmetric atomic planes (Figure 4a and 4b). For (004) plane, the full width at half maximum (FWHM) of the Ge epilayer diffraction peak is estimated to $15 \pm 4$ arcs, which is very close to the value for Ge bulk ($14 \pm 4$ arcs). The FWHM of the GaAs on Ge/PGe diffraction peak was estimated at $80 \pm 4$ arcs, which matches that of the GaAs reference epilayer grown on Ge bulk in our reactor ($76 \pm 4$ arcs) (as shown in Figure S2 of the Supporting Information). It is observed that Ge and GaAs peaks are sharp and show no significant broadening, which proves the same quality as GaAs grown on Ge bulk.

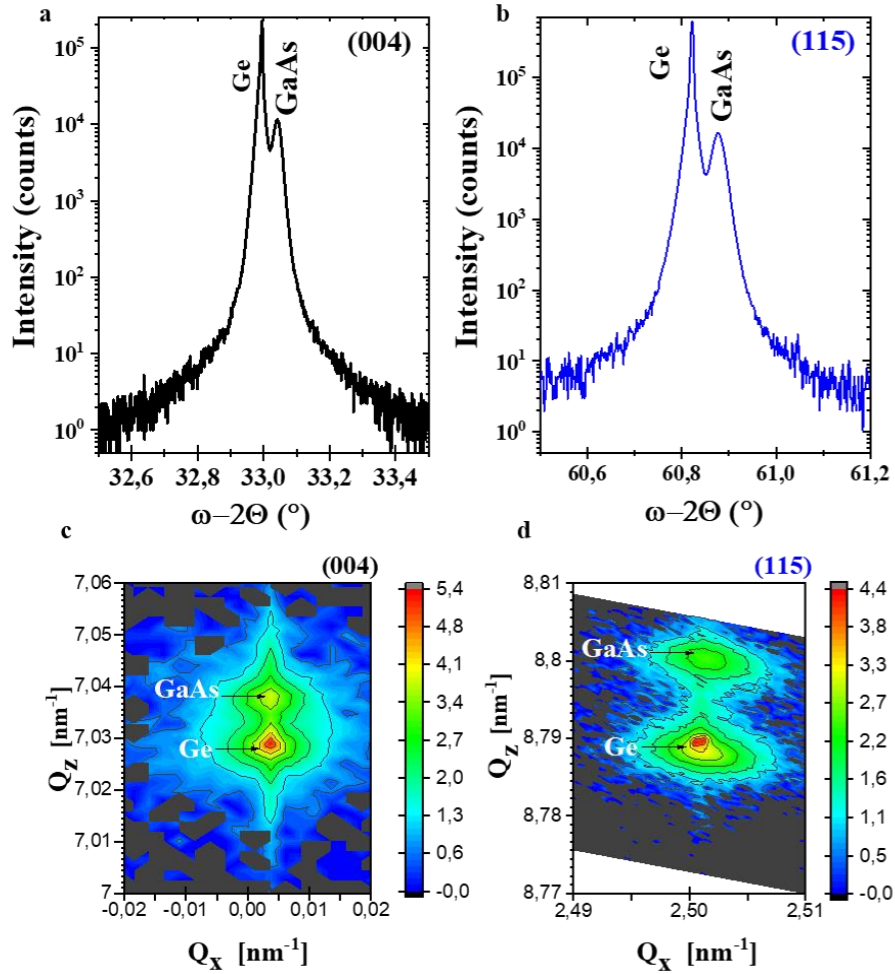

**Figure 4**. HRXRD analysis of GaAs on Ge taken around (004) (a) and (115) (b) reflections. (c) Triple axis reciprocal space (RSM) around (004) and (d) asymmetric (115).

Figure 4c and 4d display measured reciprocal space mapping (RSM) of the structure around (004) and (115) planes, respectively. Each RSM shows two distinct reciprocal lattice-point maxima corresponding to Ge and GaAs epitaxial layers, plotted as a function of their respective reciprocal space axes $Q_x$ (in-plane) and $Q_z$ (out-of-plane). For both RSM, GaAs and Ge peaks are well aligned along the vertical direction, which means that the GaAs was grown pseudo-morphically. This behavior can be explained by the fact that the deposited GaAs thickness is below the critical one,[30] leaving a fully strained layer. These results demonstrate the possibility of growing detachable high crystalline quality GaAs layers on Ge/PGe structure following the PEELER approach.

**Substrate Reconditioning and Reuse**

To assess the possibility of reusing the parent Ge wafers at the end of the process, we have designed a cost-effective reconditioning method of the Ge substrate through a chemical polishing of the surface after epitaxial layer detachment. After Ge NMs separation, the RMS surface roughness of the substrate is 20.1 nm (Figure 5a and 5c). This roughness is due to the presence of void layer residues after the separation (also observed on the SEM images in Figure S3 of the Supporting Information). This leads to the formation of inhomogeneous PGe with defects, making it unsuitable for a further epitaxial growth.[31,32]

To remove these residues, the substrate was etched with an HF-based chemical solution, with $H_2O_2$ and water. The combined use of an oxidizing agent ($H_2O_2$) and an etching agent (HF) allows etching the surface of the substrate.[33] Then, we demonstrated that a high concentration of $H_2O_2$ compared to HF and water resulted in an efficient surface flattening with σ measured at 1.3 nm (Figure 5a and 5d). The obtained value is in the same order of magnitude as that of epi-ready surface (σ at 0.3 nm Figure 5a and S1). A subsequent step is achieved with porosification of the reconditioned substrate surface, resulting in a PGe layer with σ around 2.2 nm. This roughness is comparable to that of PGe layer on epi-ready substrate (~ 2 nm Figure S1). This σ value is low-enough to ensure the suitability of the reporosified substrate for a new epitaxial growth.

Finally, after a second epitaxial growth of Ge, σ of the epitaxial layer was measured to 2.5 nm (Figure 5a and 5e)**.** This value remains low enough for a new epitaxial growth of GaAs as achieved after the first growth of Ge (σ = 0.7 nm in Figure 5a and 5b). Figure 5f shows the standard wide-angle XRD of the Ge epilayer grown on the stack PGe/reconditioned substrate. XRD patterns of the Ge bulk substrate and the Ge epilayer (1$^{st}$) growth are also included as a reference. The (002) and (004) reflections agree with those of the reference. The structural investigations rule out the presence of amorphous or polycrystalline domains and suggest the formation of monocrystalline structure. These results demonstrate a reuse sequence without loss of the surface or crystalline quality of Ge epilayer and emphasize the effectiveness of the PEELER approach.

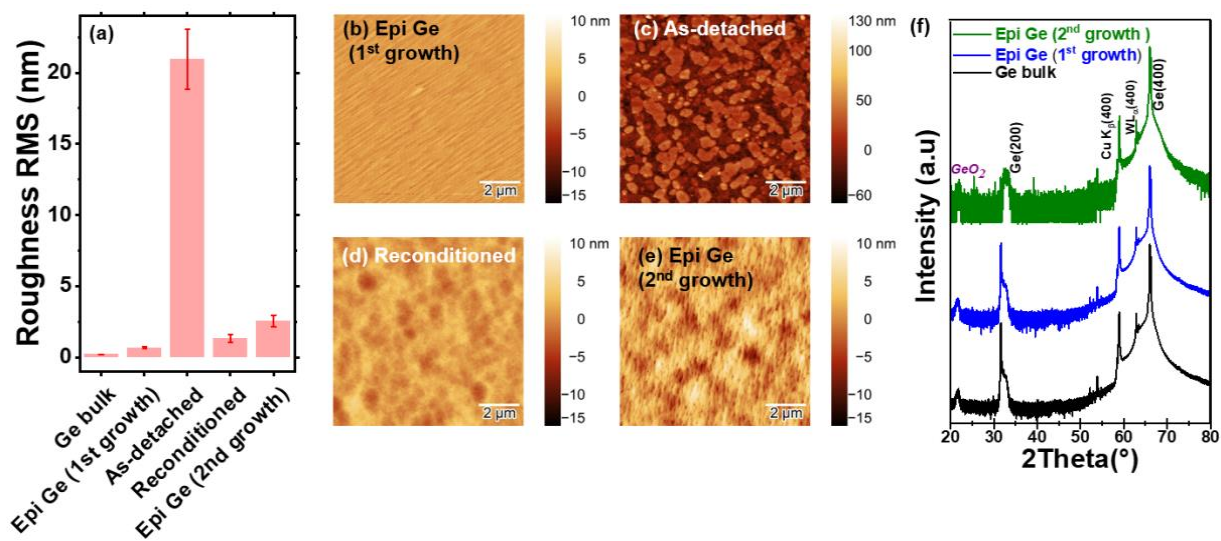

**Figure 5**. (a) RMS roughness (σ) evolution at different steps of the reconditioning process (b-e) AFM scans results (10 × 10 µm in tapping mode) of (b) after the first epi-growth, (c) as-detached substrate, (d) reconditioned substrate and (e) after epi-regrowth. (f) 2θ scan of Ge substrate, first epitaxial Ge grown on a PGe/Ge template and Epitaxial Ge regrown on PGe/Ge after reconditioning.

**Ge Consumption Analysis**

The PEELER approach significantly reduces the portion of Ge used for device fabrication compared to conventional Ge wafers[34]. As shown in Figure 6, in the case of a conventional 100 mm Ge wafer substrate (not reused), having 145 µm of thickness, 6.06 g of Ge is passed on to the device processing, e.g., solar cell production. During this manufacturing process, about 2.72 g of Ge is lost because of wafer thinning down to 80 µm[35]. On the other hand, following the PEELER approach, a substrate is expected to be reused multiple times for device manufacturing (Figure 6). Taking in consideration the growth of a 5 µm Ge layer, as it is a thickness sufficient for a Ge bottom cell,[36] only 0.21 g of Ge is consumed for a solar cell fabrication. Adding up 0.06 g of Ge lost during wastewater recycling, only 0.27 g of Ge is used per cycle. This represents more than 95% reduction of Ge consumption compared to conventional Ge wafers. Moreover, using 600 µm Ge wafers, the substrate can be reused for up to 30 times before it becomes too fragile for manipulation and is recycled. Use of thick wafers in the process limits also the loss of Ge due to the wafer sawing and wastewater treatments during the initial wafer production.[37,38]

The PEELER approach has high potential for the reduction of the substrate cost as well. According to the literature, the highest cost regarding substrate reuse rests on surface polishing, usually achieved by chemical mechanical polishing (CMP).[39] Based on our results, the surface condition of Ge NMs is directly compatible with the further growth of device structure without the necessity of CMP. Furthermore, the reconditioning was successfully demonstrated without the need for specialized equipment, e.g., CMP tools, since Ge porosification does not require supplementary polishing steps during the wafer processing. Also, thin Ge NMs do not require a back-grinding process during the cell manufacturing as they are already lightweight. As another aspect, an additional cost drop can be achievable thanks to the reduced Ge consumption as discussed previously.

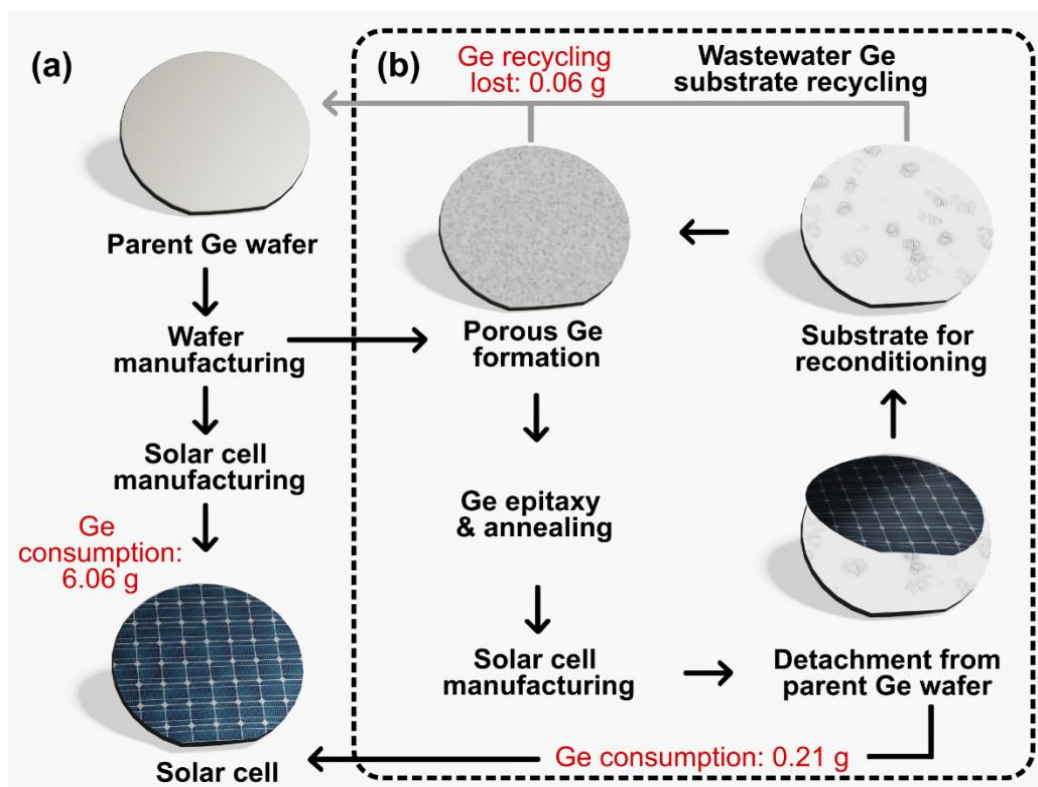

**Figure 6.** Comparison of Ge consumption between (a) the conventional Ge wafer pathway and (b) the PEELER approach illustrated in the case of solar cell fabrication.

## 3. Conclusion

In conclusion, we demonstrated the fabrication of detachable monocrystalline Ge NMs and substrate reuse, by using the PEELER approach. The proposed porosification layer separation technique uses a bipolar electrochemical process, which is demonstrated to be scalable up to 100 mm Ge wafers and allows the production of uniform PGe layers. The PGe/Ge substrate is then used to grow high-quality monocrystalline Ge NMs. The morphological transformation of PGe during the HT annealing enables the detachment of the Ge NMs. Thus, produced Ge NMs are suitable as a template for the epitaxial growth of III-V materials, an alternative to classical thick substrates. HRXRD investigations (ω-2θ and RSM) demonstrate a high crystal quality of GaAs grown on Ge NMs.

After detachment of the Ge NM, the reconditioning of the Ge substrate was achieved using a HF-based chemical solution to selectively remove PGe residues. This treatment reduces the RMS roughness of the substrate from 20 nm to 1.3 nm allowing to repeat the process on the used substrate. The PEELER process enables the production of multiple Ge NMs from a single substrate at wafer-scale. A case study of its impact on the sustainability of space solar cells production was presented, where we estimated 0.27 g of Ge consumption per PEELER cycle compared to 6.06 g consumed by conventional wafer strategy.

Finally, this approach implements a sustainable process flow by achieving a significant reduction of Ge consumption in a real case device fabrication. It implies a positive effect by avoiding unnecessary resource depletion produced by the current Ge wafer processing. Moreover, the Ge NMs take advantage of low mass and flexibility and offer a platform for a new generation of Ge and III-V based lightweight and flexible high-performance optoelectronics.

## 4. Methods

**Germanium porosification.** Porous Germanium layers were formed by bipolar electrochemical etching (BEE) in a custom-made 100 mm electrochemical cell. P-type gallium doped, 100 mm Ge wafer (provided by Umicore®) with resistivity 10–20 mΩ·cm and (100) crystal orientation with 6° miscut towards (111) was used as a substrate. Solution of HF:EtOH was used as electrolyte for the BEE. Prior to alternating BEE current, a direct a direct current was applied to induce the pore formation, as developed before.[40]

**Surface preparation.** Prior to the epitaxy growth, samples were cleaned in an acid solution of HBr:EtOH, followed by an annealing under high vacuum conditions (~$10^{-6}$ Torr).

**Epitaxial growth.** Ge and GaAs were grown in a hybrid VG Semicon V90F CBE/MBE reactor;[41,42] with a liquid nitrogen cryopanel and thermocouple for growth temperature monitoring.

The Ge was grown using solid source of Ge with K-cell heated at 1250 °C. The growth carries out in two steps with LT Ge buffer layer followed by Ge growth. The pressure, inside the chamber during the growth was ~$5 \cdot 10^{-6}$ Torr.

GaAs layer was grown using a high purity thermally cracked arsine at 950 °C and triethylgallium as arsenic and gallium sources, respectively. The growth of the 400 nm of GaAs was performed at 425 °C and ~2.7·$10^{-4}$ Torr.

**Characterization.** Cross-section images were carried out using scanning electron microscopy (LEO 1540 XB ®) to observe the porous structure before and after reorganization. An acceleration tension of 20 keV was used. The surface roughness of the porous layer and the epi-layer was measured by atomic force microscopy (AFM) with a Veeco Dimension 3000® in tapping mode, with a scan size of 10 x 10 µm$^2$.

Mapping of PGe layers was performed by ellipsometry using a J. A. Woollam Co. VASE (R) instrument (500–900 nm). 49 measurement points were measured in radial pattern with distribution of 22.5° and radial spacing of 1.25 cm.

The crystalline quality of the Ge epi-layer was evaluated by using high-resolution transmission electron microscope (HR-TEM, Talos 200X) and by X-ray diffractometer (SMARTLAB, Rigaku).

## Acknowledgments

N. Paupy, Z. O. Elhmaidi, A. Chapotot, T. Hanuš and J. Arias-Zapata contributed equally to this work. LN2 is a joint International Research Laboratory (IRL 3463) funded and co-operated in Canada by Université de Sherbrooke and in France by CNRS as well as ECL, INSA Lyon, and Université Grenoble Alpes (UGA). It is also supported by the Fonds de Recherche du Québec Nature et Technologie (FRQNT). The authors thank G. Bertrand, R. Labrecque and the 3IT's clean room staff for technical support. The authors also thank C. Andrei and B. Langelier from Mc Master University for TEM observations; the Natural Sciences and Engineering Research Council of Canada (NSERC), Innovation en énergie électrique (InnovÉÉ), Mitacs, Umicore and Saint-Augustin Canada Electric (Stace) for financial support. A. Boucherif is grateful for a Discovery grant supporting this work. A special mention is addressed to others, who have contributed to achieve the above results: Stéphanie Sauze, Arthur Dupuy, Sofiane Abdelouhab, Mohamed El-Gahouchi, Sara Taslimi, Samuel Cailleaux, Graniel Harne Abrenica, Brieuc Mével, Jérôme Deshaies, Mazarine Polydore, Quentin Reboul, Santiago Pereira, Zachary Thuotte, Antony Marcoux, Adam Dufour, Rose-Marie Pelletier, Sara Houde, Joel Dufault, Pascal Berube, Jérémy Beauchamp and Lucas Jempson.

Supporting Information

**Wafer-scale detachable monocrystalline Germanium nanomembranes for the growth of III-V materials and substrate reuse**

*Nicolas Paupy, Zakaria Oulad Elhmaidi, Alexandre Chapotot, Tadeáš Hanuš, Javier Arias-Zapata, Bouraoui Ilahi, Alexandre Heintz, Alex Brice Poungoué Mbeunmi, Roxana Arvinte, Mohammad Reza Aziziyan, Valentin Daniel, Gwenaëlle Hamon, Jérémie Chrétien, Firas Zouaghi, Ahmed Ayari, Laurie Mouchel, Jonathan Henriques, Loïc Demoulin, Thierno Mamoudou Diallo, Philippe-Olivier Provost, Hubert Pelletier, Maïté Volatier, Rufi Kurstjens, Jinyoun Cho, Guillaume Courtois, Kristof Dessein, Sébastien Arcand, Christian Dubuc, Abdelatif Jaouad, Nicolas Quaegebeur, Ryan Gosselin, Denis Machon, Richard Arès, Maxime Darnon & Abderraouf Boucherif\**

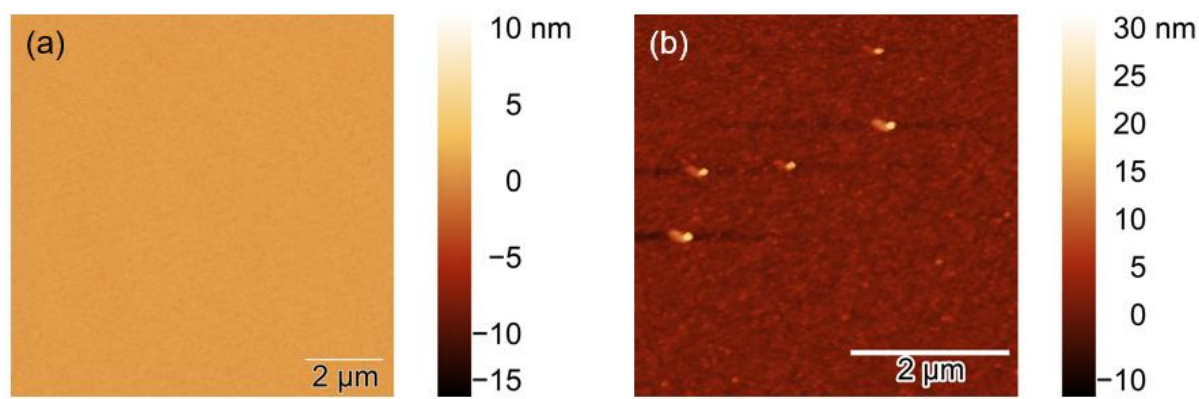

**Fig. S1.** AFM scan of (a) bulk substrate (10 µm x 10 µm) (b) the porous layer after the first cycle (5 µm x 5 µm)

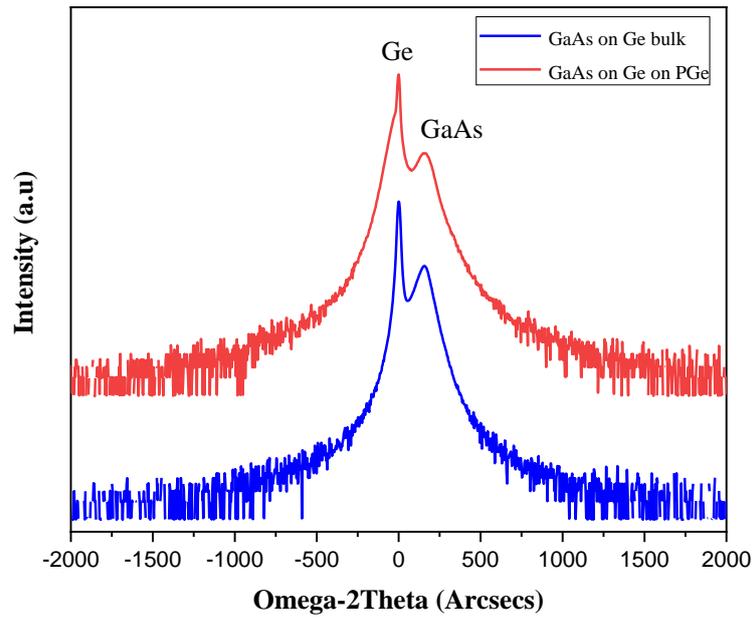

**Fig. S2.** HRXRD analysis of GaAs on Ge bulk and GaAs on Ge/PGe taken around (004)

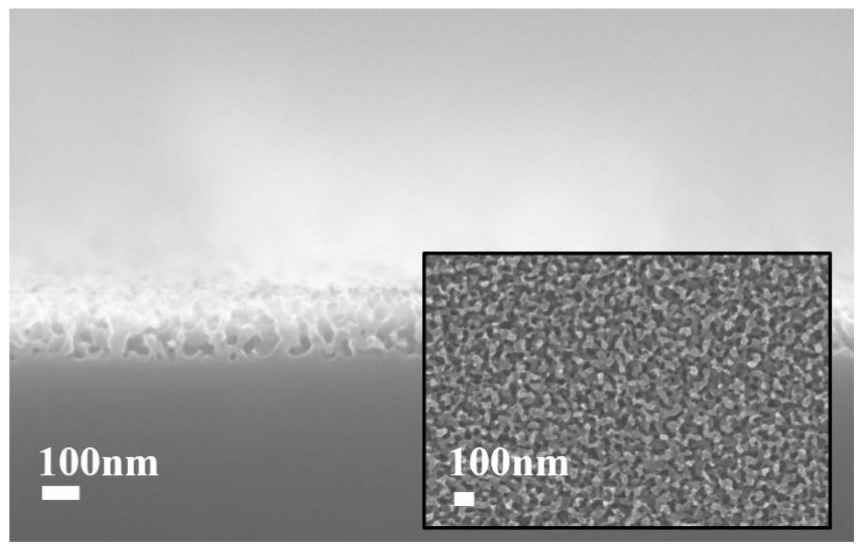

**Fig. S3.** Cross-section SEM images as-detached substrate.